# Nonlinear Optical Microscopy of Semiconductor–Metal Nanocavities

Riya Varghese,[1,*] Shambhavee Annurakshita,[1,] Yaraslau Tamashevich,[1,] Abhiroop Chellu,[2] Subhajit Bej,[1] Heikki Rekola,[3] Jari Lyytikäinen,[2] Hanna Wahl,[1] Matias Schildt,[1] Ali Panahpour,[1] Tapio Niemi,[1] Marco Ornigotti,[1] Petri Karvinen,[3] Mircea Guina,[2] Teemu Hakkarainen,[2,*] and Mikko J. Huttunen[1]

[1]Photonics Laboratory, Physics Unit, Tampere University, Korkeakoulunkatu 3, 33720, Tampere, Finland

[2]Optoelectronics Research Centre, Physics Unit, Tampere University, Korkeakoulunkatu 3, 33720, Tampere, Finland

[3]Center for Photonics Sciences, University of Eastern Finland, P.O. Box 111, FI-80101 Joensuu, Finland

[*]email: riya.varghese@tuni.fi, teemu.hakkarainen@tuni.fi

**Author notes**

*These are equal corresponding authors.

## Abstract

We use second- and third-harmonic generation (SHG and THG) microscopy to investigate the nonlinear optical response of GaAs nanocavities embedded in a gold film and compare them to bare GaAs nanocavities. Our results reveal that the surrounding metallic environment significantly modifies both the intensity and spatial distribution of the nonlinear signals. When the harmonic wavelength is spectrally detuned from the nanocavity resonance the effects due to metallic environment start suppressing the SHG contrast. Numerical simulations confirm that at a 1060 nm pump wavelength, the SHG produced at 530 nm is suppressed due to the dominant plasmonic response of gold. Meanwhile the THG produced at 353 nm which coincides with the nanocavity resonance enables high-contrast imaging. Furthermore, by shifting the pump to 710 nm, aligning SHG at 356 nm with the nanocavity resonance, we recover strong SHG contrast, demonstrating a pathway to enhanced imaging of metal–semiconductor heterostructures.

## 1. Introduction

Semiconductor nanocavities are widely used photonic structures due to their ability to confine light at subwavelength scales[1–3]. These structures enable strong light–matter interactions, making them essential for applications such as enhancing spontaneous emission[4,5], subwavelength lasers[6,7], nonlinear optics[8] and integrated photonics.[9,10] However for conventional dielectric nanocavities, the mode volume cannot be smaller than a cubic wavelength, limiting their ability to achieve extreme field confinement and enhancement.[11,12]

To overcome these limitations, hybrid metal–semiconductor nanocavities have emerged as a promising alternative, combining the advantages of high-Q dielectric nanocavities with extreme field confinement associated with metallic structures.[13,14] Embedding semiconductor nanocavities inside metallic surroundings can significantly modify the optical properties of the nanocavities. It can increase the refractive index contrast, reduce the mode volume and suppress the leaky modes leading to efficient light confinement and a strong Purcell effect.[15,16] Given the growing interest in these structures, several fabrication techniques have been developed to achieve precise control over their geometry and optical properties.[17–20] However, the diversity in fabrication methods necessitates robust characterization techniques to effectively evaluate their performance.

Electron microscopy techniques such as transmission-electron microscopy and scanning-electron microscopy (SEM) offer unparalleled spatial resolution, enabling direct imaging of nanoscale features. However, these techniques are generally invasive and require further sample preparation that may alter the properties of the studied system.[21,22] Additionally, electron-based imaging is typically limited to near-surface characterization, making it less effective for probing embedded structures.[23] On the other hand, linear optical characterization techniques like confocal[24] and fluorescence[25] microscopy enable direct visualization of structural and optical features with possibility of depth characterization. Fluorescence-based techniques can suffer from photobleaching and phototoxicity, which can degrade the sample over time, limiting their applicability for repeated measurements. In addition, it would be advantageous to investigate the properties of optical nanocavities in label-free manner,[26] instead of indirect studies based on changes in fluorescence emission. Confocal microscopy offers improved depth-resolved characterization but still face challenges in achieving high spatial resolution and signal specificity in complex nanostructured systems. Therefore, development of optical methods capable of three-dimensional (3D), depth-resolved characterization is crucial to facilitate studying nanocavities in their native environments.

Second- and third-harmonic generation (SHG and THG) are powerful nonlinear optical (NLO) processes that provide insights into structural and material properties at the nanoscale.[27] These nonlinear effects have been extensively studied in various nanostructures,[28,29] including metal–semiconductor heterostructures.[30–34] In SHG and THG processes two and three photons, respectively, interact to combine into a single photon of higher energy. Their parametric nature ensures a non-destructive approach, while the shorter generated wavelengths enhance spatial resolution, making them ideal for high-precision imaging.[27] Given these advantages, NLO imaging techniques further enhance the ability to probe such structures by providing optical sectioning and label-free contrast, making them well-suited for investigating embedded nanostructures.[35–37]

Building on these, in this work, we perform SHG and THG microscopy to investigate the optical properties of both pure cylindrical GaAs nanopillars acting as nanocavities and hybrid gold–GaAs nanostructures where the semiconductor nanocavities are formed inside gold films. We find out, that while THG microscopy can be used to investigate both kinds of nanocavities, the achieved contrast in SHG modality depends sensitively on the used input pump wavelength for the hybrid nanocavities. At 530 nm, the SHG signal exhibits poor contrast due to strong contributions from the surrounding metal film. However, longitudinal scans confirm that the cavities do contribute to the SHG signal, even though it is overshadowed by the nonlinear response from the metal. We explain these results by the changes in the optical properties of the samples and identifying the nanocavity resonances. Interestingly, contrast of SHG modality could be improved by adjusting the SHG emission wavelength to align with the nanocavity resonance, thereby improving their visibility.

## 2. Materials and Methods

### 2.1. Sample Nanofabrication

Two samples consisting of isolated yet periodically arranged GaAs nanopillars on GaAs substrate and GaAs nanopillars partially embedded on gold film were designed, fabricated, and structurally imaged via SEM. These materials were chosen because of their archetypical properties and widespread use in nanophotonics. The fabrication of samples started with molecular-beam epitaxy growth of semiconductor layers on GaAs substrates. First, a 100 nm GaAs buffer layer was grown on the substrate, followed by 50 nm of $Ga_{0.51}In_{0.49}P$ as an etch-stop layer. A 100 nm layer of GaAs was grown on top of the $Ga_{0.51}In_{0.49}P$, following which 100 nm of AlGaAs was grown. Finally, the $Al_{0.7}Ga_{0.3}As$ layer was overgrown with a 200 nm thick layer of GaAs. Electron-beam lithography was employed to pattern the top-most GaAs layer. A layer of negative resist (SX AR-N 8200, allresist GmbH) was spin-coated on the GaAs surface followed by an electron-beam exposure (Raith EBPG 5000+ESHR) in circular areas of diameter with a 2 mm period indicating where the nanopillars were intended to be located. This periodic arrangement was selected beforehand to isolate individual nanostructures for structural and nonlinear characterization. The patterned samples were then developed (developer AR 300-44, allresist GmbH) to remove the unwanted resist. Inductively coupled plasma reactive-ion etching (ICP-RIE) was employed to etch the maskless areas of the developed samples resulting in the formation of GaAs nanopillars with vertical sidewalls. The etch mask remaining on top of the nanopillars was then removed by dipping the samples in a buffered-oxide etch solution.

One of the samples was set aside as the "reference" sample containing only bare-standing GaAs nanopillars on GaAs substrate. For the rest of the paper, this sample will be called "Sample A". For the other sample ("Sample B"), the process continued with metallization of the GaAs nanopillars. Electron-beam-evaporated gold was directed at the sample installed in a downward-facing 45° angle. Conformal coating of the nanopillar sidewalls was ensured by rotating the sample at a constant speed. The angled rotational evaporation geometry meant that a nominal coverage equivalent to 1500 nm of evaporated gold was reduced by one-third, thus forming a 500 nm thick gold layer on the semiconductor surface. Note that at this stage of the process, the GaAs nanopillars were completely embedded inside a film of gold and a substrate removal procedure was performed to be able to optically access the nanopillars from the bottom side (or side in contact with the substrate). The top surface of the gold film was glued to an InP carrier while each semiconductor layer starting from the GaAs substrate up until the $Al_{0.7}Ga_{0.3}As$ layer directly below the GaAs nanopillars was sequentially removed using selective chemical etching. Following the substrate removal process, the final sample geometry consisted of an InP carrier hosting a 500 nm gold film containing arrays of GaAs nanopillars with their top faces exposed. Since the gold film thickness exceeds largely its skin depth at the relevant wavelengths used in NLO microscopy (**Section 2.2.**), the film approximates well the response of optically thick gold. SEM was performed to check the quality of the fabricated samples. The underlying substrate side of each sample was glued to a stage-fitting microscopy glass slide for subsequent nonlinear microscopy experiments.

**2.2. Nonlinear Microscopy**

Nonlinear microscopy was performed to investigate the nonlinear signals from the fabricated nanostructures. The technique is a proven way to selectively excite and spatially map the nonlinear response of individual nanostructures with a sub-micron spatial resolution.[36] For this purpose, a custom scanning-based nonlinear microscope equipped with a femtosecond laser source (wavelength output of 1060 nm, repetition rate of 80 MHz, pulse duration of 140 fs) was used. The details of the experimental setup can be found in the supplementary material (**Figure S2**).[38] After beam attenuation, linear polarization filtering, expansion and collimation, the beam was directed to a microscope objective (Nikon CFI LU Plan Fluor Epi P, numerical aperture of 0.8). This objective was used to focus the incident linearly polarized light onto the sample, which was mounted on a computer-controlled motorized stage (Mad City Labs). The back-scattered nonlinear signals from the sample were collected using the same objective and discriminated from the fundamental excitation wavelength ($\lambda_{ex}$) by appropriate dichroic, shortpass and bandpass filters (with designed wavelengths at 532 ± 9 nm and 356 ± 15 nm close to the expected SHG and THG wavelengths for a 1060 nm excitation). The filtered SHG and THG signals were then directed to separate photomultiplier tubes with identical specifications. Two-dimensional raster-scanning was performed either along the transversal (*xy*) or longitudinal (*xz* or *yz*) planes of the microscopic sample region. Before scanning, the desired region of the sample was viewed using the brightfield microscopy arm of our microscope. Unless stated, the nonlinear experiments were performed at room temperature using linear polarization along *y*, pixel dwell time of 50 ms, and pixel-scanning resolution of 0.1 μm. The input average power used while mapping was monitored before the objective and set to 5 mW. In a relevant demonstration later, the SHG from Sample B was also measured using $\lambda_{ex}$ of 710 nm tuned from the same laser source with an input power of 1.2 mW.

**3. Results and Discussion**

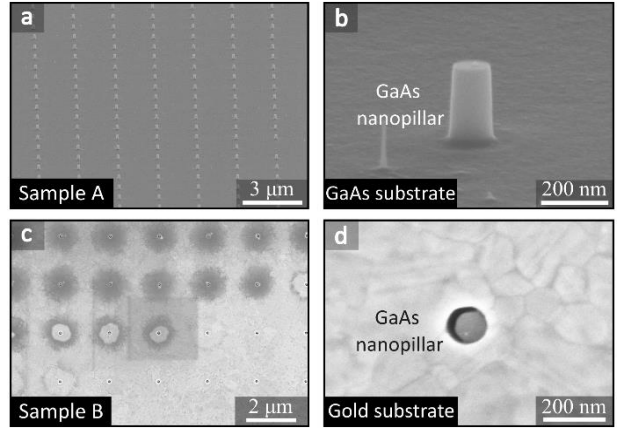

**Figure 1.** Tilted-view SEM images of Sample A: (a) GaAs nanopillar array (diameter of 115 nm, height of 200 nm, period of 2 µm) on GaAs substrate and (b) individual nanopillar on GaAs substrate. Top-view SEM images of Sample B: (c) Heterostructure array composed of GaAs nanopillars (diameter of 115 nm, height of 200 nm, period of 2 µm) embedded in gold film and (d) individual nanopillar in gold substrate.

SEM images of Sample A (**Figures 1a** and **1b**) revealed isolated nanopillars with vertical sidewalls. Structural measurements taken from SEM data of several nanopillars showed that the nanopillars exhibit an average diameter of 115 nm and height of 200 nm. Similarly, periodically arranged and isolated nanostructures that are embedded in the gold film (**Figure 1c**) are found in Sample B. Although clear structural variations are evident among the heterostructures, a representative zoom image (**Figure 1d**) showed that the exposed GaAs nanopillar top is looking symmetric while being surrounded by gold film crystallites with randomly oriented grain boundaries.

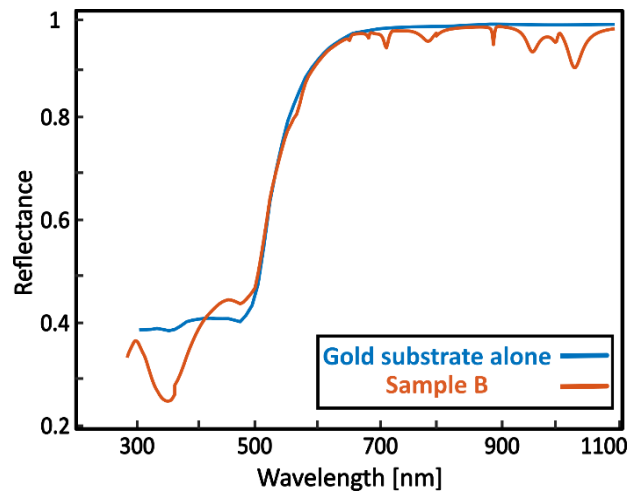

**Figure 2.** Simulated normalized reflectance curves of gold substrate alone and Sample B.

To better understand the influence of metallic embedding on the optical properties of the nanocavities, we simulated the normalized reflectance spectra for both a gold substrate alone and GaAs nanopillars embedded in a gold substrate (Sample B). The simulation was performed using COMSOL Multiphysics. The reflectance spectrum of the gold substrate alone (**Figure 2**, blue trace) shows high reflectance for wavelengths greater than 600 nm, characteristic of its metallic response. However, when GaAs nanopillars are embedded in the gold substrate (**Figure 2**, orange trace), the spectral response is significantly altered. A distinct reflectance dip appears near 350 nm, which is absent in the gold substrate alone, indicating a nanocavity resonance. This spectral feature at 350 nm implies, that contrast in NLO imaging modalities towards nanopillar structures could be enhanced by having a signal wavelength coinciding with 350 nm. In contrast, signal at 530 nm, the reflectance spectrum of Sample B closely resembles that of the bare gold substrate, implying that the optical response at this wavelength is dominated by the metallic background rather than the nanocavity resonance, resulting thus in poor imaging contrast. Based on this reasoning, we chose the excitation wavelengths that either did or did not coincide with this 350 nm resonance to enhance contrast in NLO imaging. The electromagnetic field distributions at the applied pump wavelengths ($\lambda_{ex}$ = 1060 nm and 710 nm) are also given in the supplementary material (**Figure S1 c, d**).

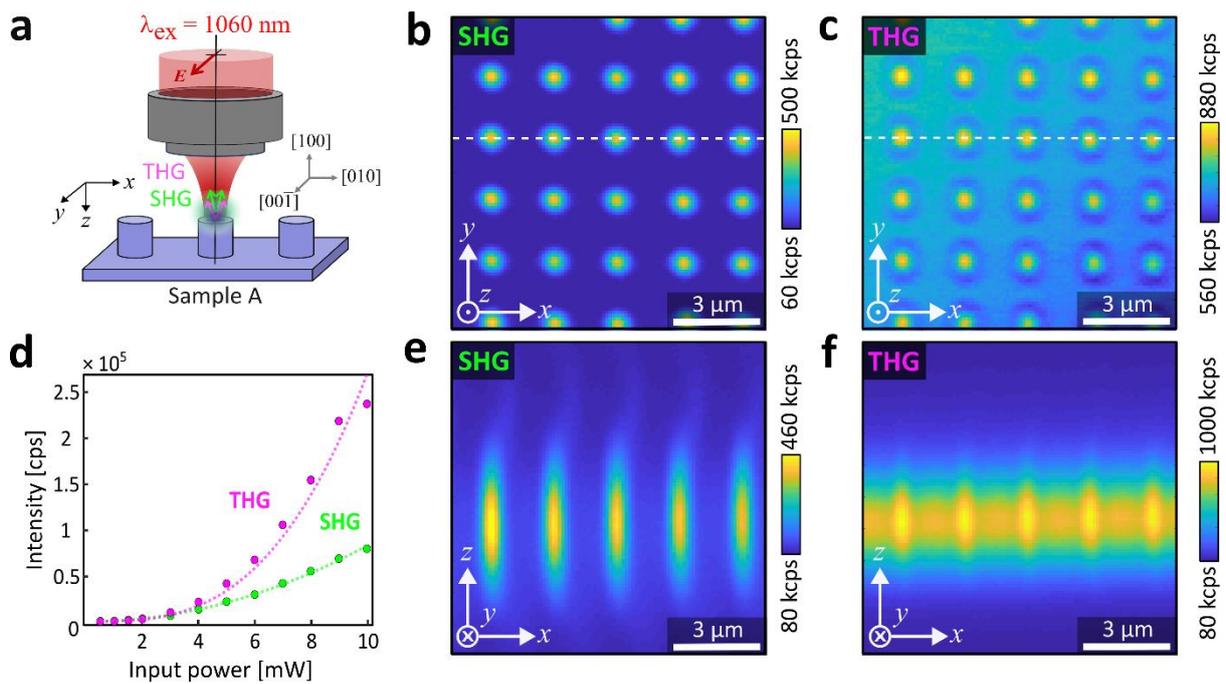

**Figure 3.** (a) Schematic of the excitation geometry for Sample A. Both the laboratory and molecular crystal frames are shown. SHG and THG are collected in the backscattering geometry with the arrows representing the THG and SHG signals collected through the objective. Transversal *xy* scanning maps of the (b) SHG and (c) THG signals from the same region of Sample A. (d) Power dependence of the acquired SHG and THG signals [in counts per second (cps)]. Experimental data are shown together with corresponding second- and third-order curve fits. Longitudinal *xz* scanning maps of the (e) SHG and (f) THG signals from the same area of the GaAs nanopillar array. These longitudinal scanning maps were acquired at the position of dashed white line in (b) and (c).

We then show our control NLO microscopy results from Sample A at the excitation wavelength $\lambda_{ex}$ of 1060 nm. It is worth noting that the generation of nonlinear signals from semiconductors, especially from GaAs, are naturally promoted by the combined effects of their crystal structure (e.g., zincblende or wurtzite), large nonlinear optical coefficients and geometric shape.[39] For zincblende GaAs, there is only one second-order nonlinear coefficient, i.e., $d = d_{14} = d_{25} = d_{36}$.[27] While the second-order response should in principle vanish under plane-wave excitation of this crystal structure, this is not strictly true for strong focusing conditions (**Figure 3a**) where the resulting focal fields contain 3D electric field components allowing cross-polarized excitations.[40,41] Repeated raster-scannings under identical illumination conditions reveal that the measured signals are mostly coming from the periodically arranged GaAs nanopillars (**Figures 3b**). Similarly, we observe a nonvanishing THG signal from the nanopillars (**Figures 3c**). The relative signal difference of the SHG (THG) from the nanopillars and the semiconductor substrate is ~4×10³ (3×10³) kcps. Note that at the air–substrate interface, the nonlinear signals are non-zero but remain lower in value than signals originating from the nanopillars. Small variations in the nonlinear signals from each nanopillar are possibly associated to imperfections.

The spatial distribution of SHG signals (along *xy*) from the nanopillar appears to be circularly symmetric with respect to its location. In contrast, the spatial distribution of the corresponding THG signals appears slightly asymmetric with measured line widths of ~0.5 μm and ~0.6 μm along the *x*- and *y*-directions. The asymmetry could be due to nearby scattering of *y*-polarization induced THG signals from nanopillars that are aligned along the *y*-axis and known polarization dependency of THG in nonlinear yet isotropic crystals.[42,43] The results verify the strongly nonlinear character of semiconductor nanostructures in agreement with previous works.[44,45] The measured nonlinear signals from the individual GaAs nanopillar at the SHG wavelength are found to follow a quadratic power dependence (**Figure 3d,** green trace) and cubic power dependence (**Figure 3d,** purple trace) at the THG wavelength. To further validate these nonlinear signals, raster-scans of the same sample region were acquired using the fixed $\lambda_{ex}$ in tandem with other bandpass filters inserted before the photomultiplier tubes. The signals are found to be significant only at the expected SHG and THG wavelength (Supplementary material, **Figure S3**). In order to investigate the spatial origin of the nonlinear signals, we leveraged the optical sectioning and the depth-resolving capabilities of our NLO microscopy setup, to perform a longitudinal *xz* scanning (**Figures 3e** and **3f**). These mappings, acquired along the dashed white line in Figures 3b and 3c, confirm that the nonlinear signals primarily originate from the nanopillars and exhibit clear contrast against the substrate.

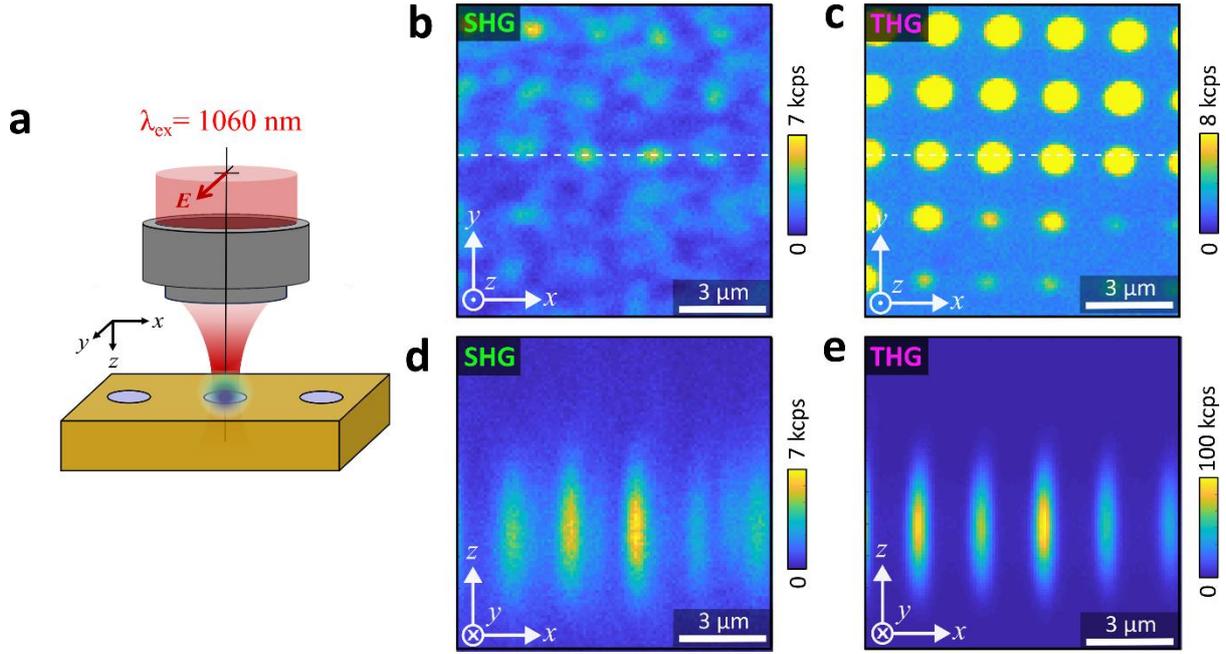

**Figure 4.** (a) Schematic of the excitation geometry for Sample B with the laboratory axis. Transversal $xy$ scanning maps of the (a) SHG and (b) THG signals from the same region of Sample B. Longitudinal $xz$ scanning maps of the (d) SHG and (e) THG signals from the same area of the GaAs nanopillar array. These longitudinal scanning maps were acquired at the position of dashed white line in (b) and (c).

We next show our NLO microscopy results from Sample B at a $\lambda_{ex}$ of 1060 nm. We first validated the nonlinear signals for Sample B using raster scans with different bandpass filters and observed that the signals were found to be significant only at the expected SHG and THG wavelengths (Supplementary material, **Figure S4**). The transversal scanning maps do not exhibit significant nonlinear signals at the SHG wavelength (**Figures 4b**). Instead, the map revealed the presence of SHG signals that are heterogeneously distributed across the scanned area. This lack of contrast in SHG maps can be explained by the simulated reflectance spectra (**Figure 2,** orange trace), which indicate that at the SHG wavelength (532 nm) the optical field is not strongly confined within the nanocavities, leading to a reduced enhancement of the SHG response and making it difficult to resolve the nanopillars in the SHG images. SHG is electric-dipole-forbidden for centrosymmetric materials like gold, so these SHG signals are possibly arising from surface effects, i.e., oriented grain boundaries of the crystalline domains present in the air–gold interface (**Figure 1d**). Without prior information about the design of Sample B and relying only on a single SHG microscopy scan at this $\lambda_{ex}$, these results alone would imply lack of GaAs nanopillars in the chosen region, suggesting failure of the fabrication of the hybrid gold–GaAs nanocavities. However, the equivalent transversal scanning maps of the THG channel revealed the presence of nonvanishing THG signals from the periodically arranged nanopillars (**Figures 4c**) indicating that in reality the quality of Sample B was not compromised. Since the THG wavelength (353 nm) coincides with the nanocavity resonance (**Figure 2**, orange trace), contrast is preserved, allowing clear identification of the nanopillars. The simulated electric field distribution of Sample B at the SHG and THG wavelength is shown in supplementary material (**Figure S1 a, b**). Due to nanopillar imperfections, slight variations in the THG maps are also expected. It is also noticeable that the gold film is showing nonzero THG signals, which is expected to occur at the air–gold substrate interface due to tight focusing conditions. These signals, however, are found to be lower than the background THG from Sample A due to the difference in the third-order susceptibility values of GaAs and gold.[27] However, we also note that the substrate of Sample A is composed of other optical materials aside from GaAs, which can potentially affect the nonlinear signals emanating from the substrate.

The longitudinal scanning maps (**Figure 4d**), however, showed that the SHG signals are found to be still present at the nanopillar locations, despite being masked by the stronger response from the gold substrate in the transverse maps. This highlights the importance of depth-resolved scanning in distinguishing contributions from different structural components. However, the overall SHG intensity remains low, as the excitation wavelength does not coincide with the nanocavity resonance, limiting the enhancement of the nonlinear response. In contrast, the longitudinal THG scan clearly differentiates the nanopillars from the gold substrate, consistent with the transverse maps. The THG signals are significantly stronger than the SHG signals, as the THG wavelength falls within the nanocavity resonance, leading to enhanced nonlinear response. However, this distinct contrast in THG response was not as evident in the transverse scans, further emphasizing the role of depth scanning in accurately capturing the optical behavior of the system.

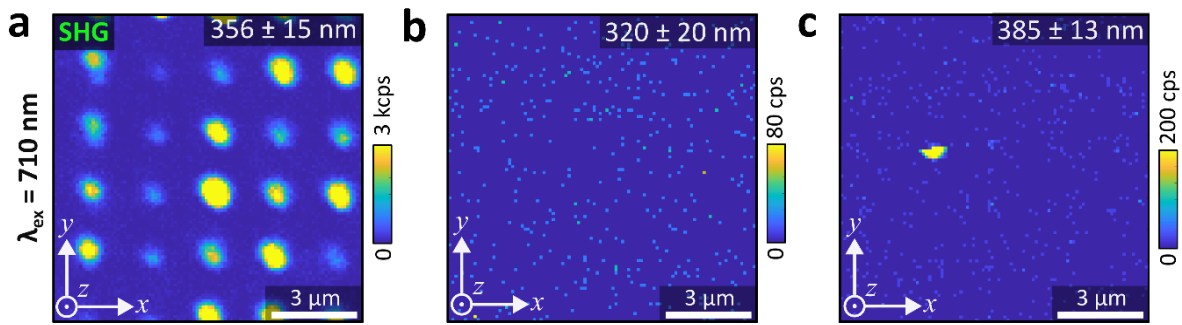

**Figure 5.** a) Schematic of the excitation geometry for Sample B using $\lambda_{ex}$ of 710 nm. Transversal *xy* scanning maps of the sample at (b) SHG (356 ± 15 nm) wavelengths and using different bandpass filters with central wavelengths of (c) 320 ± 20 nm and (d) 385 ± 13 nm. Asymmetry in the SHG maps is attributed to slight optical misalignment of the scanning beam.

The simulation results showed that the linear response of Sample B (**Figure 2,** blue trace) exhibit a nanocavity resonance at 350 nm (see E-field distributions in the Supplementary material **Figure S1 a**), near to the THG channel of our nonlinear microscope. To further investigate the role of the nanocavity resonance in suppressing SHG contrast, we performed NLO microscopy of Sample B using $\lambda_{ex}$ of 710 nm (**Figure 5a**) inducing SHG near 355 nm at the resonance of Sample B. As a result, the influence of the metallic environment surrounding the cavity is minimized, allowing for enhanced contrast in our SHG images. The resonant excitation now revealed the presence of SHG emission from GaAs nanopillars that are now distinct from the background (**Figures 5b** and **5c**). The expected THG near 236 nm, on the other hand, was not verified due to lack of UV optics and detection capabilities. While all experimental parameters were set to be the same as in the first experiment with pump near 1060 nm, the input power used was set to 1.2 mW. Future investigations could focus on polarization-resolved measurements, including circular polarization effects, to further explore the symmetry properties of the system.

## 4. Conclusion

We investigated the NLO response of metal–semiconductor nanocavities using SHG and THG microscopy. We have fabricated heterostructures that are composed of GaAs nanopillars embedded in gold film and a corresponding control sample made of GaAs nanopillars on GaAs substrate. Using NLO microscopy, we have found drastic changes in the strengths and spatial distributions of the SHG and THG signal intensities from the individual heterostructure and the control indicating that the surrounding environment of the nanopillar can considerably affect their linear and nonlinear optical responses. Our results show that the presence of a metallic coating introduces strong localization effects that, depending on the used pump wavelength, can suppress the SHG emission and achievable imaging contrast. The lack of contrast was particularly visible when the SHG occurred at 530 nm wavelength, being far away from the nanocavity resonance. Numerical simulations confirmed that at a pump wavelength of 1060 nm, the SHG wavelength (530 nm) was spectrally away from the nanocavity resonance, leading to suppressed SHG contrast due to a strong nonlinear response of the gold layer. In contrast, at the THG wavelength (353 nm), which coincided with the nanocavity resonance, we observed good contrast in the THG images, enabling clear visualization of the nanopillar structures even when embedded inside a highly reflecting gold film. Interestingly, by shifting the pump wavelength to 710 nm, with respective shift in the SHG signal to 355 nm and thus aligning with the nanocavity resonance, an enhanced SHG image contrast for visualizing gold-embedded nanocavities was restored. These results provide broader insight into the characterization of photonic nanocavities beyond semiconductor-based systems, demonstrating how nonlinear optical imaging techniques can be used to study a wide range of nanocavities. Furthermore, our findings shed light on SHG and THG generation in ultra-small mode volume semiconductor structures, where strong field confinement has the potential to enhance nonlinear processes. As these imaging techniques require no additional sample preparation, they hold promise for exploring the optical responses of individual nanocavities in diverse material platforms.

**Acknowledgements**

A.C., S.B., J.L., M.O., P.K., A.P., M.G., T.H. and M.J.H. acknowledge PREIN and Research Council of Finland for financial support. The following foundations are acknowledged for doctoral research support: Jenny and Antti Wihuri Foundation (S.A.), Finnish Cultural Foundation (Y.T and S.A.), Tekniikan Edistämissäätiö (R.V.). We acknowledge Godofredo Bautista for his contributions toward this work.

## Author Contributions

T.H. and Godofredo Bautista conceived the idea. J.L. carried out molecular beam epitaxy growth of the samples. M.G. supervised work related to molecular beam epitaxy. Work of A.P., R.V., S.A., and M.S. was supervised by M.J.H. and Godofredo Bautista. H.R. and P.K. performed electron beam lithography patterning. S.B. and H.W. carried out dry etching of nanopillars. S.B. and A.C. were involved in metallization of the nanopillars. A.C. and H.W. performed SEM characterization of the nanopillars. T.H. and T.N. supervised nanofabrication of the samples. R.V., S.A. and M.S. performed the nonlinear optics experiments, curated, and validated the data. R.V. and S.A. processed the data. R.V., S.A., A.P. and M.J.H analysed the nonlinear microscopy data. R.V., Y.T., M.O., A.P., T.H. and M.J.H designed the numerical experiments. R.V., A.P. and Y.T. implemented the numerical experiments. R.V., Y.T, M.O., A.P., T.H. and M.J.H. analysed numerical results. All co-authors contributed to writing the manuscript.

## Conflict of Interest

The authors declare no conflict of interest.

## Data Availability Statement

The data that support the findings of this study are available from the corresponding authors upon reasonable request.

## Keywords

# Supplementary Info

## Electric field distribution

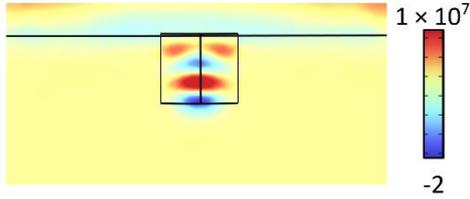
(a) $E_x$ field at 350 nm

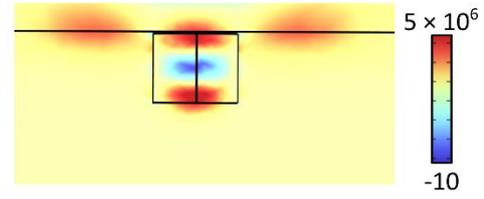
(b) $E_x$ field at 530 nm

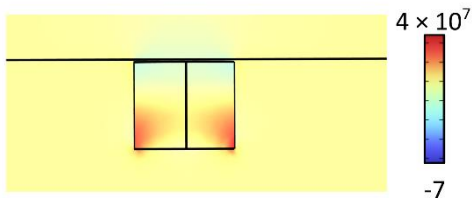
(c) $E_x$ field at 1060 nm

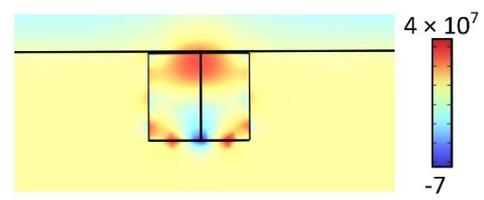
(d) $E_x$ field at 710 nm

Figure S1: Simulated x-component of electric field distribution ($E_x$) in the xz plane for gold embedded nanopillars (a) at 350 nm (THG wavelength for 1060 nm excitation) (b) Electric field at 530 nm (SHG wavelength for 1060 nm excitation) (c) at 1060 nm fundamental pump wavelength, (d) at 710 nm pump wavelength.

The numerical simulations were performed using COMSOL Multiphysics with the Wave Optics Module. The gold-embedded nanopillar structures were modeled using experimental material parameters, incorporating wavelength-dependent permittivity for gold (Johnson and Christy) and GaAs. A plane wave excitation was used with electric field polarized along the x-axis and the periodicity of the nanopillars were 2μm. The perfectly matched layers were used at the top and bottom part of the simulation domain and periodic boundary conditions were applied along the lateral boundaries.

## Nonlinear microscopy setup

A schematic diagram of the setup is shown in Figure S2.

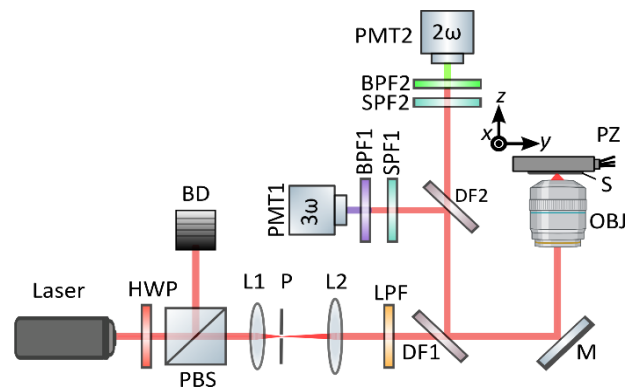

Figure S2: Schematic diagram of the NLO microscopy setup. The microscope is powered by a laser source (1060 nm, 80 MHz, 140 fs,) and have the following components: half-wave plate (HWP), polarizing beamsplitter (PBS), beam dump (BD), lenses (L1, L2), pinhole (P), longpass filter (LPF), dichroic filters (DF1, DF2), mirror (M), microscope objective (OBJ), sample plane (S), piezo-scanners

(PZ), shortpass filters (SP1, SP2) relevant for THG and SHG wavelength, bandpass filters (BPF1, BPF2) relevant for THG and SHG wavelength and photomultiplier tubes (PMT1, PMT2).

## Validation of nonlinear results

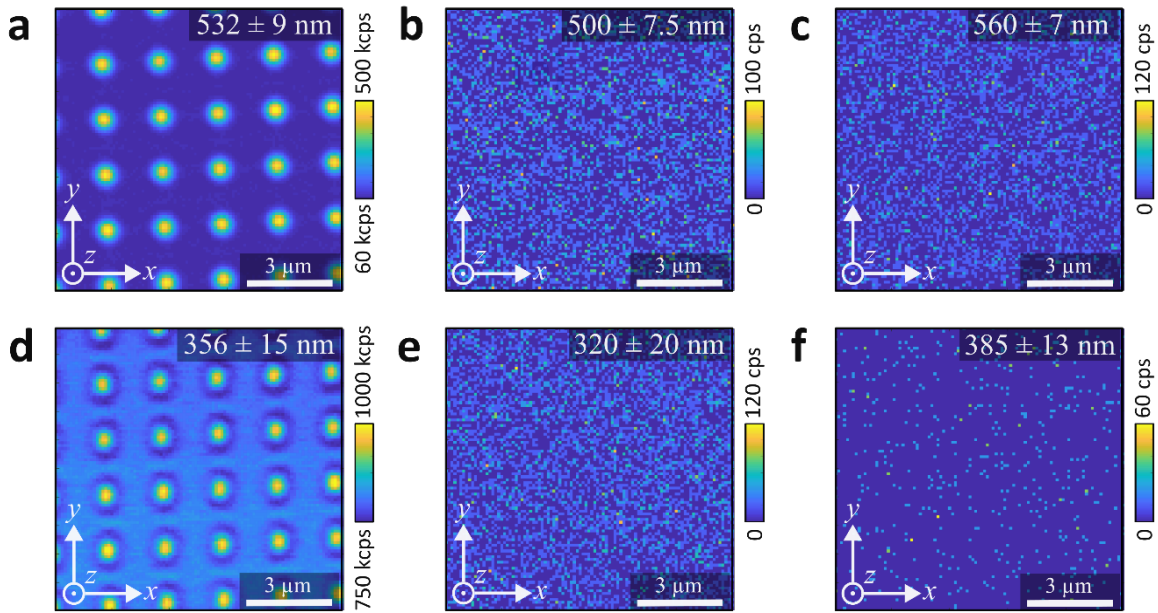

Figure S3: Transversal *xy* scanning maps of the (a) SHG and (d) THG signals of Sample A using 1060 nm excitation. Repeated scanning maps of the same area in (a,d) using bandpass filters with central wavelengths of (b) 500 ± 7.5 nm), (c) 560 ± 7 nm, (e) 320 ± 20 nm and (f) 385 ± 13 nm.

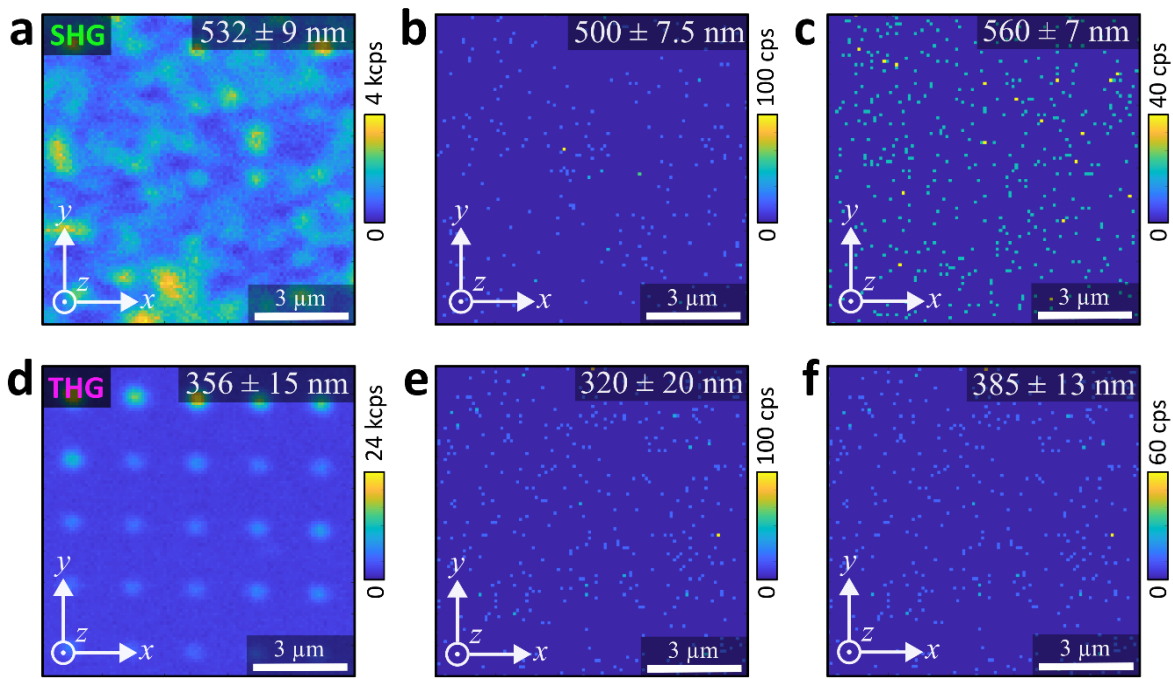

Figure S4: Transversal *xy* scanning maps of the (a) SHG and (d) THG signals of Sample B using 1060 nm excitation. Repeated scanning maps of the same area in (a,d) using bandpass filters with central wavelengths of (b) 500 ± 7.5 nm), (c) 560 ± 7 nm, (e) 320 ± 20 nm and (f) 385 ± 13 nm.

For verification, additional bandpass filters were applied to confirm the spectral integrity of the nonlinear signals. Specifically, for SHG at 532 nm, filters at 500 nm and 560 nm were used. While for THG at 352 nm, filters at 320 nm and 385 nm were applied. The same verification procedure was followed when using a 710 nm pump, with SHG signals at 356 nm confirmed using bandpass filters at 320 nm and 385 nm. It is visible from the results that the nonlinear signals are only visible at the relevant wavelengths (Figure S3 a, d and Figure S4 a, d).